\documentclass{PoS}
\usepackage{subfigure}
\usepackage{wrapfig}

\newcommand{\ie}{{\it i.e.}}
\newcommand{\eg}{{\it e.g.}}

\newcommand{\cf}[1]{{Fig.~\ref{#1}}}
\newcommand{\ct}[1]{{Table~(\ref{#1})}}

\title{Associated-quarkonium production}

\ShortTitle{Associated-quarkonium production}

\author{\speaker{Jean-Philippe Lansberg}\\
        IPN Orsay, Paris Sud U., CNRS/IN2P3, F-91406 Orsay, France\\
        E-mail: \email{Jean-Philippe.Lansberg@in2p3.fr}}

\abstract{We discuss the growing interest to measure associated-quarkonium production in a number of channels at the LHC. 
Whereas back-to-back production of quarkonium + isolated photon provides a unique way to extract gluon TMDs, 
observables such as quarkonium + $W$/$Z$ can be of great help to better understand the quarkonium production mechanism 
as well as to shed light on double-parton scatterings. Along these lines, we also argue that 
quarkonium-pair production is a potentially rich source of information which only has  started to be harvested.
Finally, we discuss the relevance of studying the production of quarkonium + heavy-quark, as \eg~$J/\psi$ + charm and $\Upsilon$ + non-prompt $J/\psi$.}

\FullConference{XXII. International Workshop on Deep-Inelastic Scattering and Related Subjects,\\
		28 April - 2 May 2014\\
		Warsaw, Poland}

\begin{document}

\section{Introduction}
\vspace*{-5pt}

Since the start-up of the LHC, the ATLAS, ALICE, CMS and LHCb detectors have collected 
data at higher energies, at higher transverse momenta, with better precision and 
with more exclusivity towards direct production compared to that which was achieved before. 
Unfortunately, all this seems insufficient to clear up the complexity of the quarkonium-production mechanism. 
In this context, a growing attention was directed towards the study of Associated-Quarkonium Production (AQP) channels. In a number of cases, 
they are expected to put specific constraints on certain parameters of the theoretical approaches proposed
to describe quarkonium production.

One of the reasons for the complications in describing these elementary reactions is certainly connected
to the large expected impact of perturbative corrections in $\alpha_s$. It is for instance well known 
that $\alpha^4_s$ and $\alpha^5_s$ corrections to the colour-singlet 
mechanism (CSM)~\cite{CSM_hadron} have a significant impact on the $P_T$ dependence of the $J/\psi$ and $\Upsilon$ cross sections observed in 
high-energy hadron collisions~\cite{Campbell:2007ws,Artoisenet:2007xi,Gong:2008sn,Artoisenet:2008fc}
as well as on their polarisation~\cite{Gong:2008sn,Lansberg:2010vq,Li:2008ym,Lansberg:2009db}.
NLO corrections to the colour-octet mechanism (COM) cannot also be overlooked since
they significantly affect polarisation predictions and the extraction~\cite{Butenschoen:2012px} of the non-perturbative octet matrix 
elements (also referred to as LDMEs). All this renders the phenomenological analyses rather complex to interpret owing to the 
large uncertainties in the current theoretical predictions. On the contrary, 
when one focuses on the $P_T$-integrated yields, both for the charmonia and the bottomonia,
the CSM contributions agree relatively well with the existing data at colliders
energies~\cite{Brodsky:2009cf,Lansberg:2010cn}. 

The introduction of new observables, such as AQP channels, may thus be of great help. 
For some, the impact of QCD corrections is expected to be smaller, thus with smaller uncertainties 
on the renormalisation scale for instance. Others are expected to be specifically discrimant towards the separation
of CO and CS contributions, being particularly sensitive to either of these contributions. Finally, some 
AQP reactions have similar properties as the inclusive-production reactions 
and simply provide further constraints for the fits of NRQCD LDMEs.

\vspace*{-5pt}
\section{Quarkonium plus Quarkonium}
\vspace*{-5pt}

We start with the first AQP ever measured, \ie~that of a pair of $J/\psi$'s. 
It has been measured for the first time  at large $x_F$
by the CERN-NA3 collaboration~\cite{Badier:1982ae,Badier:1985ri}  in the eighties. 
The rates were higher than 
expected and they seemed to be only explainable  by the coalescence of 
a double intrinsic-charm pair in the proton projectile~\cite{Vogt:1995tf}.
Recently, LHCb has also studied this process at the LHC~\cite{Aaij:2011yc}. 
It is important to realise that the cross section measured by LHCb covers a totally different region than NA3, 
which a priori should be accounted for by the conventional pQCD approaches, \eg~by the 
CSM. As a matter of fact, the $P_T$-integrated rate seen by LHCb is in very good agreement 
with the recent theoretical expectations from the CSM at LO~\cite{Qiao:2009kg,Berezhnoy:2011xy}. 
As far as the $P_T$-integrated yields are concerned, it is reasonable to say that the CSM predictions are as 
satisfactory for double  $J/\psi$ production as for single $J/\psi$ production.

Last year, we evaluated the leading-$P_T$ contribution at NLO, dubbed as NLO$^\star$, for $J/\psi$ 
pair production along with that of $J/\psi + \eta_c$ at LO~\cite{Lansberg:2013qka} using the
automated matrix element and event generator {\sc HELAC-Onia}~\cite{Shao:2012iz}. Our partial NLO$^\star$ evaluation 
has been confirmed by a full NLO evaluation by Sun and Chao~\cite{Sun:2014gca}. We have found that the NLO$^\star$ is indeed
enhanced w.r.t. LO for increasing $P_T$. It is already 8 times larger for $P_T \gtrsim$ 5~GeV and nearly 400 
times larger for $P_T \gtrsim$ 30~GeV. This is compatible with a relative suppression scaling as $P_T^{-2}$ between
the LO and NLO topologies, which justifies the use of the  NLO$^\star$ approximation (see~\cite{Artoisenet:2008fc}). 
As regards the $P_T$ spectrum for $J/\psi + \eta_c$ at LO, it lies exactly in between these cases.

\begin{figure}[hbt!]
  \begin{center} 
\subfigure{\includegraphics[width=7.5cm,clip=true]{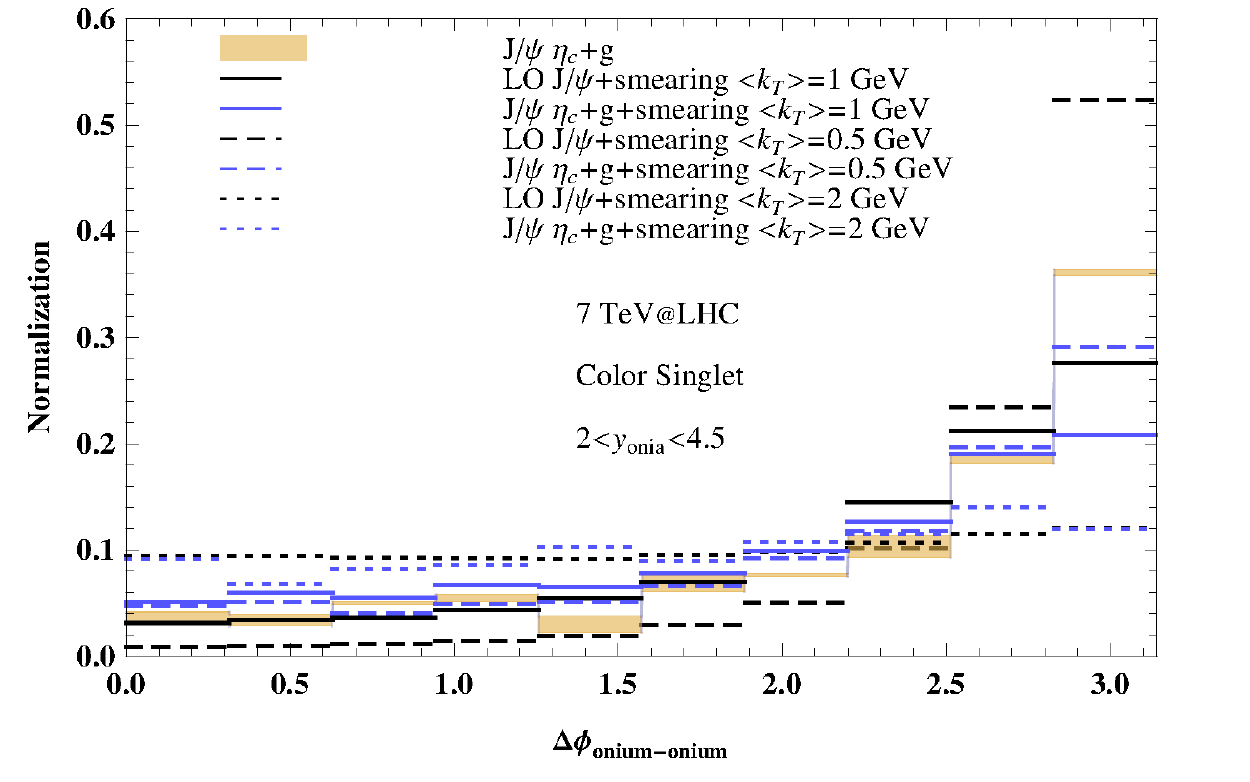}}
\subfigure{\includegraphics[width=7.5cm,clip=true]{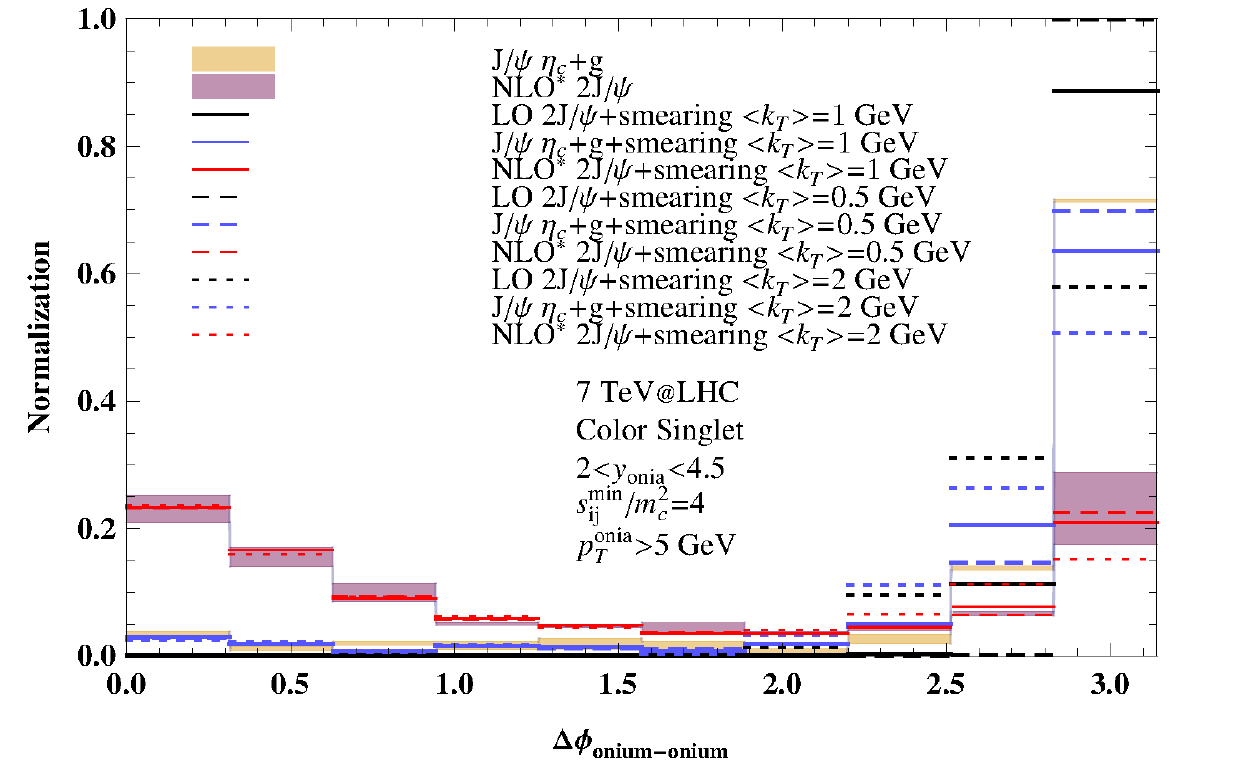}}
\caption{Azimuthal distribution of quarkonium pairs for different initial $k_T$ of the colliding partons with (left) and 
without (right) a $P_T$ cut on one of the quarkonium.
\label{fig:deltaphi}}\vspace*{-10pt}
\end{center}
\end{figure}

In addition to the yields and their kinematical dependences, we have also looked at the azimuthal correlations 
between the quarkonia. At  NLO$^{(\star)}$, the presence of a gluon in the final state of the leading topologies 
naturally creates an imbalance between the quarkonia, which are then not necessarily created back to back ($\Delta \phi = \pi$). The presence of initial $k_T$ in the colliding partons is also another source of imbalance. 
If $\langle k_T \rangle$ is as high as 2 GeV, the azimuthal distribution is completely flattened 
(see \cf{fig:deltaphi} (left)); in other words, 
the quarkonia are completely decorrelated in azimuth. If one imposes a $P_T$ cut on one of the quarkonia, they
remain produced back to back, except for $J/\psi-J/\psi$ at NLO (see \cf{fig:deltaphi} (right)). Indeed, there is a possibility that the $J/\psi$ pair be
produced recoiling on a gluon. In such a case, the quarkonia are ``near'' to each other ($\Delta \phi \simeq 0$).
These results, in any case, show that one has to be careful when trying to separate out the contribution of Single-Parton Scatterings (SPS) 
 --expected to be {\it anti}-correlated-- from that of Double-Parton Scatterings
--expected to be uncorrelated--  uniquely from the analysis of the azimuthal distributions. Our findings
confirm the discussion  in~\cite{Kom:2011bd}.

Finally, let us add that double $J/\psi$ production has recently been  studied by D0 at Fermilab~\cite{Abazov:2014qba} 
and by CMS at the LHC~\cite{Khachatryan:2014iia}. These studies respectively brought in new information about the
rapidity and the $P_T$ difference dependence. A possible next step might be to look at the polarisation as we did at NLO$^{(\star)}$ 
in~\cite{Lansberg:2013qka}.



\vspace*{-5pt}
\section{Quarkonium plus heavy quarks: $J/\psi + c$ and $\Upsilon + \hbox{non-prompt } J/\psi$}
\vspace*{-5pt}

In addition to quarkonium-pair production, LHCb has recently investigated  the production of $J/\psi$ with charm hadrons~\cite{Aaij:2012dz}.
Depending on the scheme considered for the number of flavours, this process can come from partonic sub-processes such as $gc \to J/\psi c$~\cite{Brodsky:2009cf} 
or $gg \to J/\psi c \bar c$~\cite{Artoisenet:2007xi}.   A comparison with the 
LHCb data~\cite{Aaij:2012dz} is not straighforward since one has to take into account in the kinematics the charm-quark fragmentation. 
In addition, the LHCb data are only for $P_T^C$ of the charm hadrons above 3 GeV. In $2\to 2$ processes, this automatically excludes $P_T^{J/\psi} \leq 3$~GeV. 
QCD corrections may therefore be important for these kinematical configurations. As for now, these reactions have only been evaluated at LO and 
the current theoretical uncertainties from the scales and from the charm-quark mass are unfortunately large.
In spite of this, the theory-data comparison hints at an important DPS contribution. Yet, since the
$P_T$ spectra of the $J/\psi$'s produced with a charmed hadron and of those without tend to differ, SPS contributions may be sizeable. 
Overall, much still has to be learnt from these interesting pieces of data.

\begin{wrapfigure}{r}{0.375\textwidth}
  \vspace{-20pt}
  \begin{center}
\subfigure[Leading-$P_T$ CSM and COM graphs]{\includegraphics[width=5cm,clip=true]{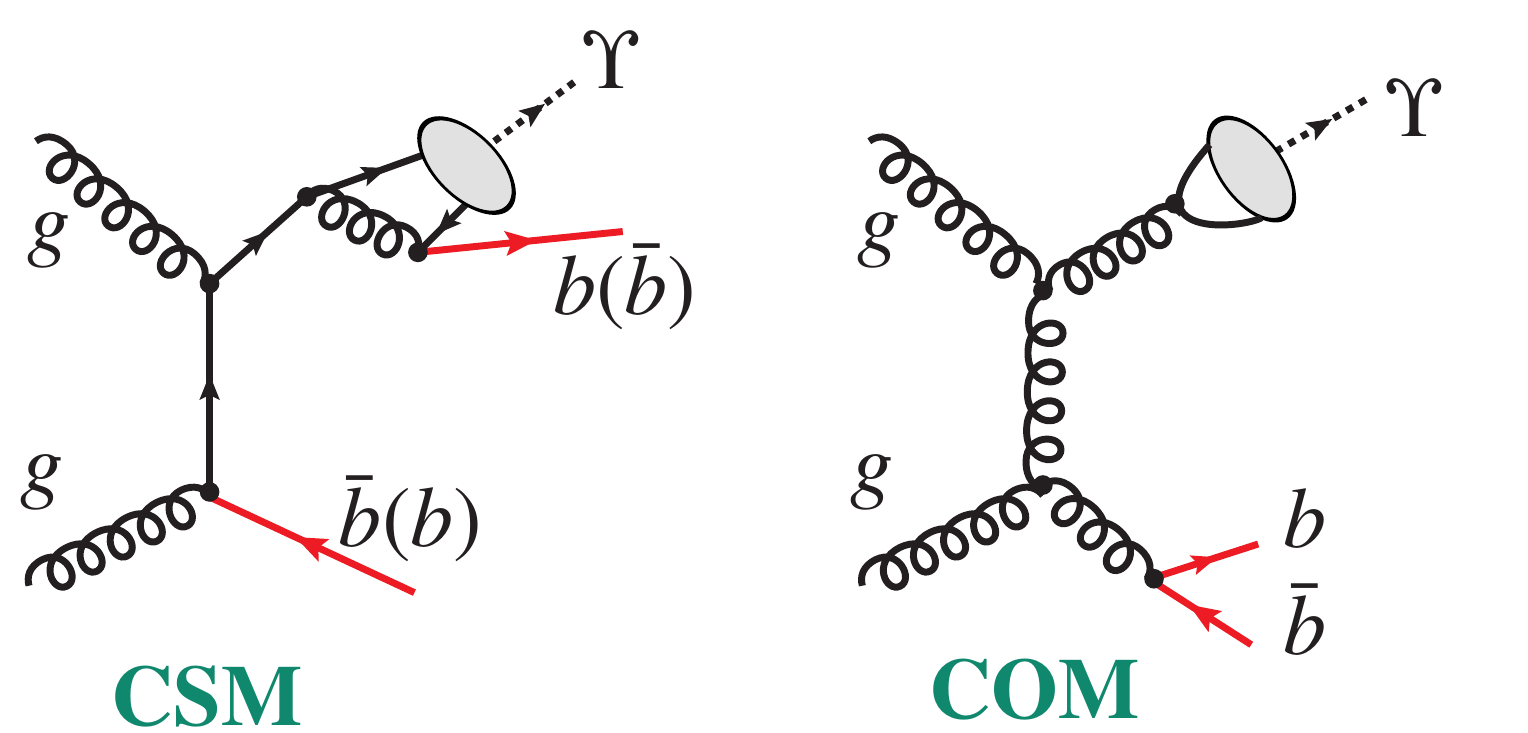}}
\subfigure[CSM crosss-section at the LHC]{\includegraphics[width=5.3cm,clip=true]{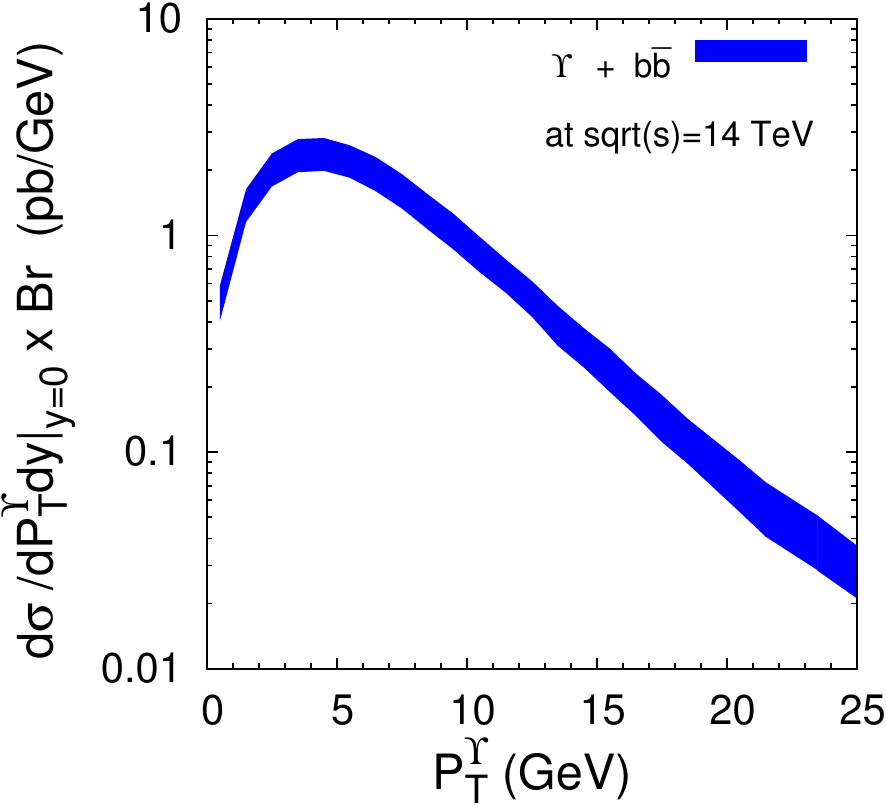}}
  \end{center}
  \vspace{-20pt}
  \caption{Hadroproduction of $\Upsilon + b\bar b $}\label{fig:upsilon_plus_b}
  \vspace{-10pt}
\end{wrapfigure}

Along the same lines, bottomonia can also be produced with a beauty hadron. Since these are usually studied via their
weak decay into a $J/\psi$, final states such as $\Upsilon + \hbox{non-prompt } J/\psi$ are also worth investigating. As for 
$J/\psi\,+\,$charm, one can look at the event topology, in particular, at how ``near'' (or ``far'') the heavy-flavoured particle is (w.r.t. the quarkonium).
This can indeed be an interesting way to pin down the dominant production (SPS) mechanism\footnote{In the case of DPS production, 
one naturally expects the absence of any kind of correlations.}. If one assumes that such a reaction is initated by gluon fusion, the 
leading-$P_T$ processes for, respectively the CSM and COM contributions, are as depicted on \cf{fig:upsilon_plus_b} (a). One clearly sees that the 
COM contributions can only produce two beauty hadrons (\ie~$J/\psi$'s) back-to-back to the $\Upsilon$. At large enough $P^\Upsilon_T$, one would not expect at all
beauty hadrons ``near'' the $\Upsilon$. This is precisely the opposite of what one expects from the CSM, where one beauty hadron would be ``near'' and 
one ``away''. The cross section for the LHC kinematics is shown on \cf{fig:upsilon_plus_b} (b). Since about 1\% of $b$ quarks
eventually end up decaying into a $J/\psi$, one can reasonably expect up to 400 $\Upsilon + \hbox{non-prompt } J/\psi$ events at $P^\Upsilon_T\simeq 10$~GeV 
with an integrated luminosity of 20 fb$^{-1}$ at the LHC.

\vspace*{-5pt}
\section{Quarkonium plus a back-to-back isolated photon}
\vspace*{-5pt}

The production of a quarkonium associated with an isolated photon has been discussed 
in the literature at many instances since the early 90's. The yields have been evaluated at NLO in~\cite{Li:2008ym}
and a partial NNLO (NNLO$^\star$) evaluation has been performed~\cite{Lansberg:2009db}. This
observable is an interesting complement to inclusive measurements. For instance, 
it may put stringent constraints on the CO LDMEs\footnote{An updated NLO analysis 
taking into account CO channels recently showed that some of the NLO fits 
of CO LDMEs~\cite{Butenschoen:2012px} provided unphysical predictions --to be precise, negative yields-- for $J/\psi+\gamma$ production~\cite{Li:2014ava}.}. Yet, 
the expected rates are necessarily lower than for inclusive quarkonium production and the theoretical
uncertainties are not necessarily smaller. 

In this context, it is worth noting that the requirement for back-to-back quarkonium--photon production selects out
an interesting part of the phase space where (i) neither the QCD corrections, (ii) nor the CO contributions, are kinematically
 enhanced, and where (iii) an extension of collinear factorisation -- the TMD factorisation-- can 
fully  be applied, providing a rigourous set of tools to study the transverse dynamics of the gluon content of the proton.
This is what we discussed in~\cite{Dunnen:2014eta}. 

In particular, we computed the expected rates at the LHC, which are summarised in \ct{tab:xsec-onium-gamma}. 
On the one hand, we can see that, at the LHC, this observable is essentially from gluon fusion, 
making it an excellent gluon probe. On the other hand, 
it is most likely a pure CS yield in the $\Upsilon$ case whereas, in the $J/\psi$ case, the predicted 
CS yield is above the CO ones where the cross section for accessible events 
($J/\psi-\gamma$ invariant mass pair between 20 and 25 GeV) is the highest. For the sake of 
TMD factorisation applicability, 
it may also be worth isolating the $J/\psi$ in order to prevent any contamination from CO transitions.

\begin{wraptable}{r}{0.67\textwidth}
  \vspace{-30pt}
  \begin{center}
\includegraphics[width=10cm,clip=true]{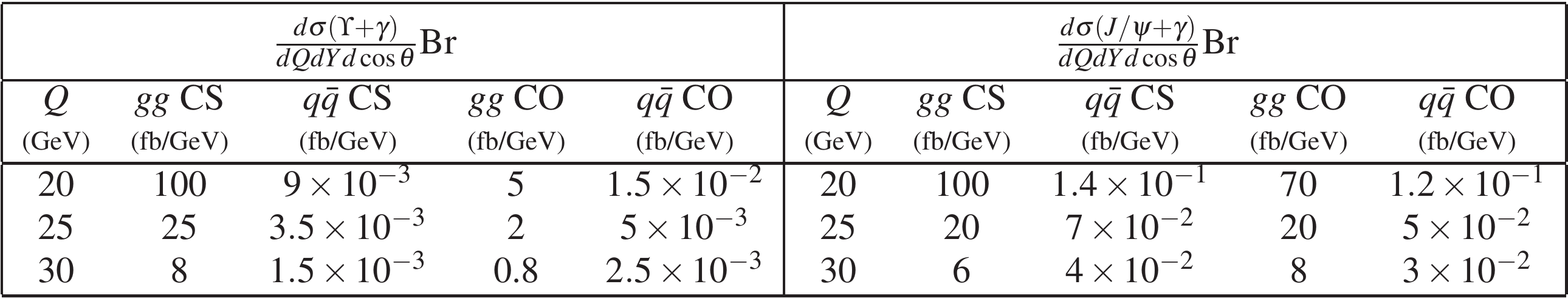}
  \end{center}
  \vspace{-10pt}
  \caption{ Cross section for  back-to-back quarkonium--photon production at $\sqrt{s}=$7~TeV for 
gluon fusion and quark-annhilation and for CS and CO transitions for different
pair invariant masses, $Q$, and for $|Y|<0.5$ \& $|\cos\theta|<0.45$ (see \protect\cite{Dunnen:2014eta}).}\label{tab:xsec-onium-gamma}
  \vspace{-10pt}
\end{wraptable}

Our conclusion is that such an observable can already be studied at the LHC with 
data on tapes (about 20 fb$^{-1}$ of $pp$ collisions). In particular, one should be 
able to extract -- for the first time -- the transverse momentum dependence of the gluon content in the proton and
to tell whether the distribution of linearly polarised gluon in an unpolarised proton is nonzero.
It has to be noted that such an observable is also sensitive to gluons at lower energies, such as at the proposed
fixed-target experiment using the LHC beams~\cite{Brodsky:2012vg} (AFTER@LHC). With 20 fb$^{-1}$ of $pp$ data at $\sqrt{s}=115$~GeV, it
should be possible to probe the gluon TMDs up to $x\simeq 0.5$~\cite{transversity} with this process at $Q\simeq 10$ GeV.

\vspace*{-5pt}
\section{Quarkonium plus $W$/$Z$ bosons}
\vspace*{-5pt}

The production of a quarkonium associated with a vector boson has also been the object of a number
of studies. Previous theoretical analyses of $J/\psi+W$ production at hadron colliders~\cite{Barger:1995vx,Kniehl:2002wd,Li:2010hc}
focused on the leading contributions in $\alpha$ or $\alpha_s$, only arising from COM. For a long time, it was believed that 
{\it ``$\psi+W$ offers a clean test of the colour-octet contributions''}~\cite{Barger:1995vx} 
and that {\it ``If the $J/\psi+W$ production is really detected, it would be a solid basis for testing the color-octet 
mechanism of the NRQCD''}~\cite{Li:2010hc}, the latter statement being made following a 
NLO study in $\alpha_s$. 

In~\cite{Lansberg:2013wva}, we have shown that the CSM contributions to direct $J/\psi+W^\pm$  
are not small at all as compared to that from CO transitions. These CSM contributions are due to 
two sub-processes: a) the fusion of a gluon and a strange quark which turns into a charm quark by the emission of the $W$, 
followed by the fragmentation of the charm quark into a $J/\psi+c$ pair; b) the quark $q$ and antiquark $\bar q'$
annihilation  into an off-shell photon, $\gamma^\star$, and a $W$, followed by 
the fluctuation of the $\gamma^\star$ into a $J/\psi$. The former process appears at $\alpha_s^3 \alpha$ and the latter at $\alpha^3$ compared to $\alpha_s^2 \alpha$ for the COM process. The CSM contributions were earlier disregarded since
formally appearing at higher orders in $\alpha$ or $\alpha_s$ despite the suppression of the CO in the $v$ expansion of NRQCD.

At the Tevatron, we found that the COM contribution was significantly larger than that of the CSM via $sg$ fusion, 
but of similar size as the CSM contribution via $\gamma^\star$. At LHC energies, our results was that the three contributions were of the same order. 
As a result, the total CSM cross section is about twice as large as the COM one at 7 TeV, probably a little more at
 14 TeV and at large $P_T$ --because of the most favourable running of $\alpha$ w.r.t $\alpha_s$ for increasing scales.
Yet, at LHC energies, based on the ATLAS results~\cite{Aad:2014rua}, it seems that there is a significant DPS contribution.

We have also studied the hadroproduction of  $J/\psi+Z$ and  $\Upsilon+Z$ at NLO in~\cite{Gong:2012ah}. We found 
out that the quarkonium polarisation is not affected by the QCD corrections, whereas these significantly affect the 
yield for increasing $P_T$.

\vspace*{-5pt}
\section{Conclusion}
\vspace*{-5pt}

A number of associated-quarkonium-production channels have been theoretically 
studied since the early nineties. Thanks to the large luminosities recently collected at both
Fermilab and the LHC, a number of experimental studies have been --and are being-- carried out. There is no doubt that the confrontation
between these theory predictions and these measurements will bring in very soon new information which will be very useful  in order to solve the 
quarkonium-production puzzles. They may also play a key role in the understanding of DPS physics.

\end{document}